\documentclass[prl,twocolumn, showpacs,amsmath,amssymb]{revtex4}

\usepackage{float}
\usepackage{graphicx}
\usepackage{subfigure}

\renewcommand{\vec}[1]{\mathbf{#1}} 
\newcommand{\mathcommand}[3][0]{\newcommand{#2}[#1]{\ensuremath{#3}}}
\newcommand{\ts}[2]{{#1}_{\text{#2}}} 
\newcommand{\ham}[1]{\ensuremath{\mathcal{H}_{\text{#1}}}} 
\mathcommand{\psie}{\ts{\psi}{e}} 
\mathcommand{\gamman}{\ts{\gamma}{n}} 
\mathcommand{\hypsplit}{\Delta E_{\text{HF}, \text{P}^0}} 
\mathcommand[1]{\ket}{\left| #1 \right\rangle}  
\mathcommand[1]{\bra}{\left\langle #1 \right|}  
\mathcommand{\si}{^{29}\text{Si}}
\mathcommand{\natsi}{^{28}\text{Si}}
\mathcommand{\p}{^{31}\text{P}}
\mathcommand{\mub}{\ts{\mu}{B}}  
\mathcommand{\kb}{\ts{k}{B}}     
\mathcommand{\MHz}{\text{~MHz}}
\mathcommand{\ns}{\text{~ns}}
\mathcommand{\percc}{\text{~cm}^{-3}}
\mathcommand{\cm}{\text{~cm}}
\newcommand{\transit}{\ham{N}}
\newcommand{\startstate}{{\text{P}^+\!\!\uparrow\Downarrow}}
\newcommand{\middlestate}{{\text{P}^0\!\!\uparrow\Downarrow}}
\newcommand{\badfinalstate}{{\text{P}^0\!\!\downarrow\Uparrow}}
\newcommand{\goodfinalstate}{{\text{P}^0\!\!\uparrow\Downarrow}}
\newcommand{\nonstate}{{\text{P}^0\!\!\downarrow\Downarrow}}
\newcommand{\firstorderup}{V^{(1)}_{\text{P}^+\uparrow\Downarrow\rightarrow\text{P}^0\uparrow\Downarrow}}
\newcommand{\firstorderdown}{V^{(1)}_{\text{P}^+\downarrow\Downarrow\rightarrow\text{P}^0\downarrow\Downarrow}}

\newcommand{\secondorder}{V^{(2)}_{\text{P}^+\uparrow\Downarrow\rightarrow\text{P}^0\downarrow\Uparrow}}

\newcommand{\energydiff}{E_{\text{P}^0\downarrow\Uparrow}-E_{\text{P}^0\uparrow\Downarrow}}

\begin{document}

\title{Optical Detection of a Single Nuclear Spin}
\author{Kai-Mei C. Fu}
\email{kaimeifu@stanford.edu}
\author{Thaddeus D. Ladd}
\author{Charles Santori}
\altaffiliation[Also at ]{Research Center for Advanced Science and
Technology, University of Tokyo, 4-6-1 Komaba, Meguro-ku, Tokyo
153-8904, Japan.}
\author{Yoshihisa Yamamoto}
\altaffiliation[Also at ]{NTT Basic Research Laboratories, 3-1
Morinosato-Wakamiya Atsugi, Kanagawa, 243-0198, Japan.}
\affiliation{Quantum Entanglement Project, ICORP, JST, Edward L.
    Ginzton Laboratory, Stanford University, Stanford, California
    94305-4085, USA}
\date{\today}

\begin{abstract}

We propose a method to optically detect the spin state of a
$^{31}$P nucleus embedded in a $^{28}$Si matrix.  The
nuclear-electron hyperfine splitting of the $^{31}$P neutral-donor
ground state can be resolved via a direct frequency discrimination
measurement of the $^{31}$P-bound exciton photoluminescence using
single photon detectors. The measurement time is expected to be
shorter than the lifetime of the nuclear spin at 4~K and 10~T.

\end{abstract}

\maketitle

The spin qubit embedded in a crystalline host is an attractive
choice for solid-state quantum computation due to its long
coherence time. The inevitable cost of the spin qubit's isolation
from the environment is the difficulty of measuring its quantum
state. Scanning tunnelling microscopy and single molecule
spectroscopy have pushed close to this limit by reporting single
nuclear and single electron spin
resonances~\cite{Manassen00a,Wrachtrup93a, Wrachtrup97a}, yet
measurement of the quantum state of a single spin in the solid
state remains elusive.

Nuclei in semiconductors, in particular silicon, have generated
particular interest for solid-state quantum computation. The
proposal of Kane~\cite{Kane98a,Kane00a}, in which qubits are
encoded as the spin state of single embedded \p\ nuclei, proposes
to solve the single nuclear measurement problem by adiabatically
transferring the nuclear state to the spin of the electron bound
to the \p\ impurity and measuring the electron's spin-correlated
charge state with a single electron transistor (SET).  In
practice, however, charge fluctuations produced by the SET couple
back to the \p\ nucleus via the electron's strong hyperfine
interaction, leading to a decoherence source that is not present
in bulk silicon. Magnetic resonance force microscopy has
approached single-electron-spin sensitivity, but experimental
results thus far have shown that this probe induces spin
relaxation more quickly than the needed measurement
time~\cite{Stipe01a}, casting doubts on proposals that seek its
use for quantum computation~\cite{Berman01a}.   The use of
ensembles of \si\ nuclei in bulk silicon with no metallic gates
has been proposed~\cite{Ladd02a}, but this scheme replaces the
measurement problem with the challenge of achieving high nuclear
polarization in order to maintain scalability.  In this Letter, we
propose an all-optical method to determine the nuclear spin state
of an isolated \p\ impurity in bulk silicon.  This method's
incorporation into silicon-based quantum computation architectures
could ease these difficulties.

In semiconductors, free excitons can be bound to donor and
acceptor impurity sites, forming bound excitons.  When a bound
exciton (BE) decays radiatively to a neutral impurity state, its
linewidth is characteristically narrow due to the localization of
the exciton~\cite{Karaiskaj01a}. If the hyperfine interaction
between the impurity nucleus and either the neutral impurity bound
carrier or the impurity BE is sufficiently strong, the nuclear
state of the impurity can be determined via the BE
photoluminescence (PL) energy. The particular case of the $^{31}$P
donor in a $^{28}$Si matrix is treated below.

\begin{figure}
\includegraphics[height = 55mm, keepaspectratio]{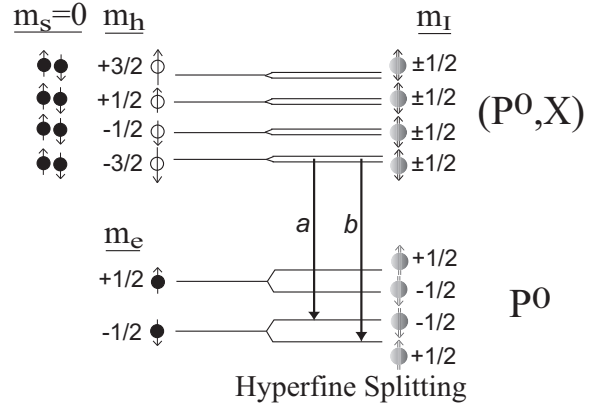}
 \caption{Energy diagram for the neutral donor
(P$^0$) and its bound exciton (P$^0$,X) in a magnetic field. The
(P$^0$,X) state is populated via capture of a resonantly excited
free exciton.  The $^{31}$P nuclear state can be determined by the
energy difference of $a$ and $b$.} \label{energy}
\end{figure}

In a magnetic field, the ground state of the $^{31}$P BE, notated
(P$^0$,X), is split into four hole Z\'eeman levels.   The ground
state of the neutral donor, P$^0$, is split into two electron
Z\'eeman levels~\cite{Ziemelis82a}.  These levels are illustrated
in Fig.~\ref{energy}.
%
%
%
%
We assume a large applied magnetic field such that the
electron/hole Z\'{e}eman interaction is much greater than the
hyperfine interaction.

The P$^0$ state is described by effective mass theory (EMT), where
the total wavefunction $\psie(\vec{r})$ of the donor electron is
given by the product of the EMT envelope function $F(\vec{r})$ and
the Bloch wavefunction $\varphi(\vec{r})= u(\vec{r})
e^{i\vec{k}\cdot\vec{r}}$.  Both the Bloch and envelope functions
are $s$-like, so the dominant part of the hyperfine interaction is
the Fermi contact term~\cite{Abragam61a},
\begin{equation}
\ham{C} = -\frac{\mu_0}{4\pi}g_0\mub\hbar\gamman\frac{8\pi}{3}\vec{I}\cdot\vec{S} \ \delta(\vec{r}),\label{contact}
\end{equation}
where $g_0$ is the free-electron $g$-factor, $\gamman$ is the
$^{31}$P gyromagnetic ratio, $\vec{I}$ is the nuclear spin
operator, and $\vec{S}$ is the electron spin operator. When an
exciton is bound to the neutral-donor site, the two electrons form
an antisymmetric spin singlet state
\cite{Kirczenow77a,Dean67a}.  Consequently, the hyperfine splitting of the
BE is determined only by the spin of the bound hole.  Since the
hole Bloch function is $p$-like, the Fermi contact term is
negligible, and assuming that the envelope function is $s$-like,
the orbital and dipolar terms will be much smaller than the P$^0$
contact hyperfine splitting.  Thus, the energy difference between
the transitions $a$ and $b$ of the (P$^0$,X) PL, shown in Fig.~1,
is determined entirely by the hyperfine splitting of the P$^0$
state. From Eq.~(\ref{contact}), this splitting is
\begin{equation}
    \hypsplit =
    \frac{\mu_0}{3}g_0\mub\gamman\hbar|\psie(0)|^2.
\end{equation}
The electron density at the $^{31}$P ($\vec{r}=0$) site has been
determined via electron spin resonance to be
$0.44\times10^{24}\percc,$ corresponding to a $\hypsplit$ of
$60\MHz$~\cite{Feher56a}. This splitting has also been calculated
with reasonable accuracy using a corrected envelope function to
account for the discrepancy in the observed and EMT ionization
energies~\cite{Kohn57a, Feher59a}.

The (P$^0$,X) state in Si decays primarily via a non-radiative
Auger process with a lifetime of 300~ns~\cite{Schmid77a}. However,
there exists a zero-phonon radiative channel with a lifetime of
2~ms~\cite{Schmid77a}. The PL linewidth of a single $^{31}$P donor
impurity is thus expected to be approximately $3\MHz$, which is
much smaller than the hyperfine splitting of 60~MHz.
Experimentally, the PL linewidth from an ensemble of $^{31}$P
impurities was measured to be less than 150~MHz at 2~K, which
includes an inhomogeneous broadening effect and was limited by the
spectrometer resolution~\cite{Karaiskaj01a}.

After Auger recombination, the electron/hole recombination energy
is imparted to the second electron, which is ionized.  To ensure a
fast recapture process of the donor electron, one can optically
excite conduction electrons. Simultaneously, one can resonantly
excite free excitons which will be captured at the neutral-donor
site. The BE will relax thermally within its lifetime
\cite{Ziemelis82a}, and in a magnetic field of 10~T at 4~K, the probability
for it to occupy the lowest Z\'{e}eman level is 0.8. Thus,
approximately 400~photons/s are emitted at the desired transitions
$a$ and $b$ in Fig.~\ref{energy}.

\begin{figure}
\subfigure[]{\includegraphics[height = 35mm,
keepaspectratio]{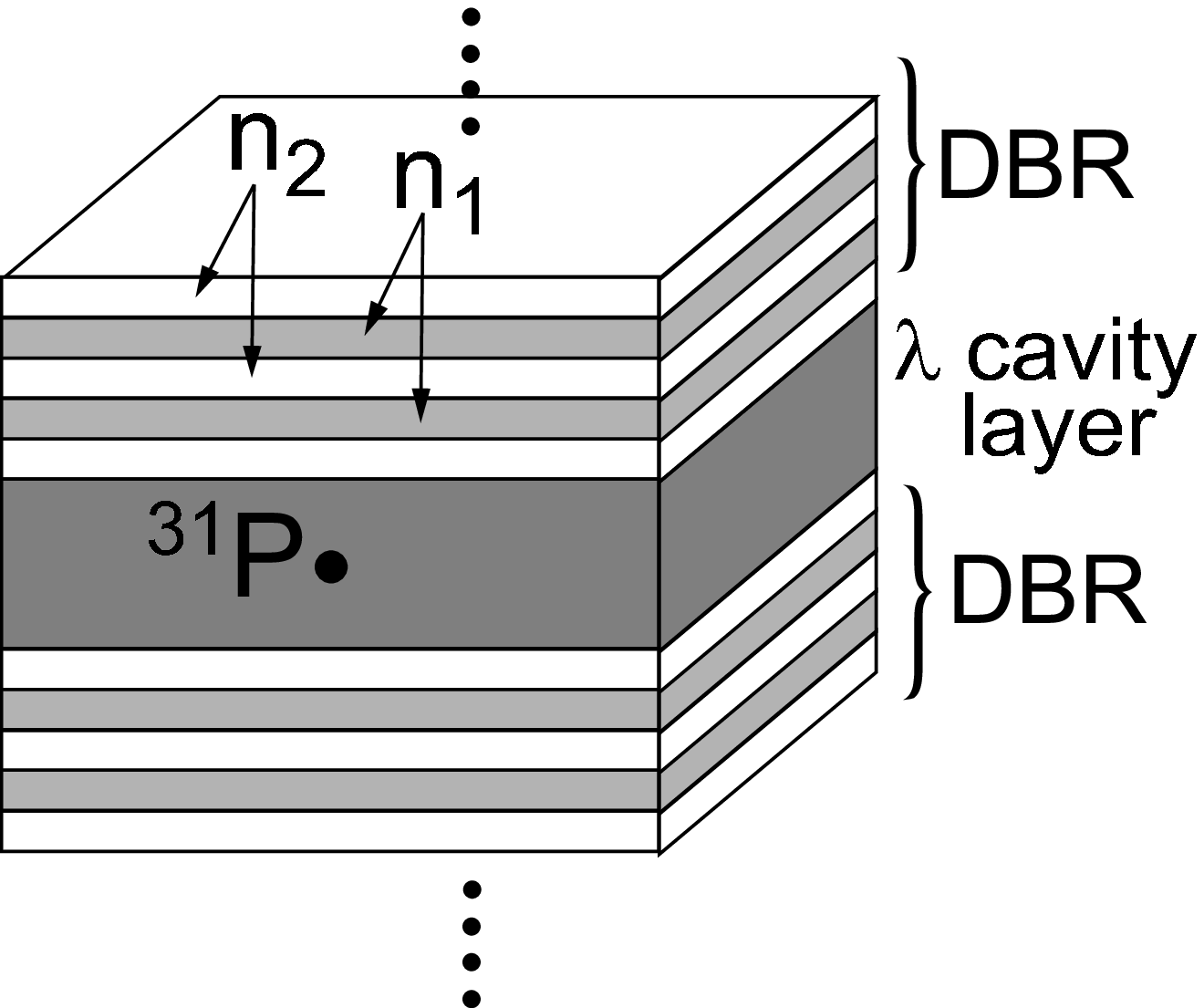}} \quad \subfigure[]{
\includegraphics[height = 33mm,  keepaspectratio]{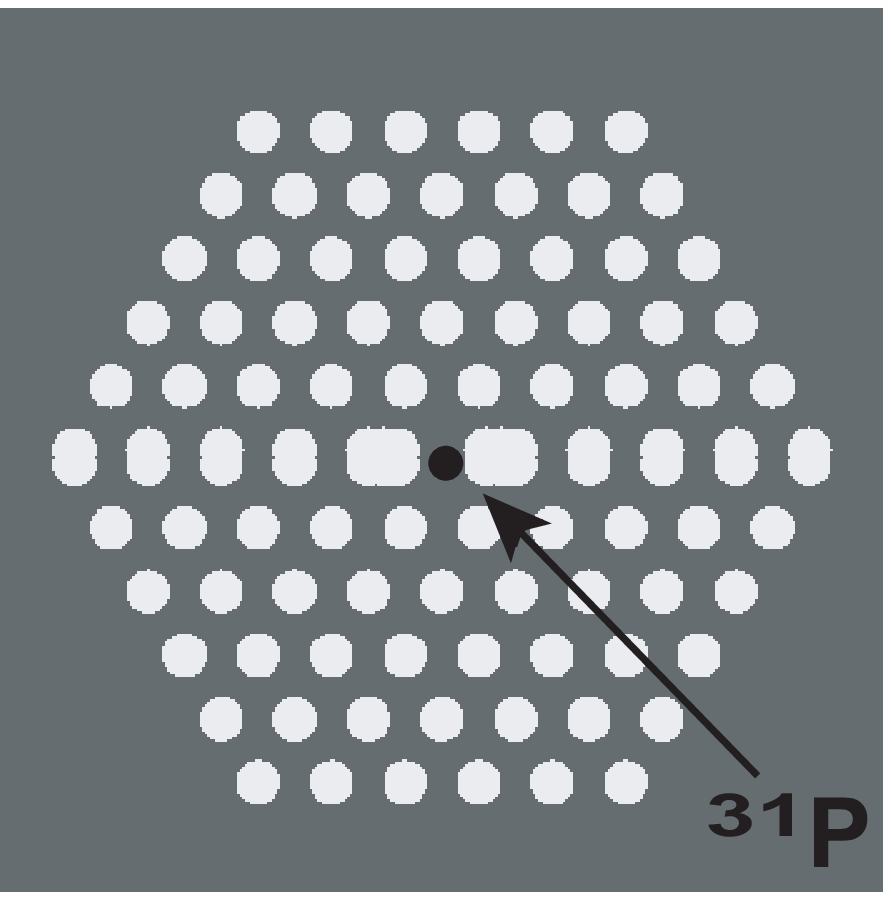}}  \caption{(a) DBR planar cavity including a
$^{31}$P impurity at the center of the one-wavelength-thick
optical cavity layer. (b) 2D photonic crystal structure including
a $^{31}$P impurity at the central defect~\cite{Vuckovic03a}.}
\label{cavity}
\end{figure}

The extraction of an emitted photon out of the
high-refractive-index Si substrate as a well-collimated beam for
optical detection is not trivial.  For this purpose, one can
incorporate a planar distributed Bragg reflector (DBR) cavity, as
shown in Fig.~\ref{cavity}(a),
%
%
%
%
at the center of which a $^{31}$P impurity is embedded.  The
cavity modifies the radiation pattern and concentrates the emitted
power in the normal direction.  With high and low DBR refractive
indices $n_1 = 3.0$ and $n_2 = 1.5$, the output coupling
efficiency ($\beta$-factor) into a beam emitted in a normal
direction can be as high as 0.8 for a random dipole
orientation~\cite{Yamamoto91a}.  This efficiency is achieved at
the cost of a decreased radiative decay rate by a factor of
3~\cite{Yamamoto91a}. The overall number of PL photons available
for optical detection would then be $400\times0.8/3 \sim
100$~photons/s.

Alternatively, a two-dimensional photonic crystal structure
(Fig.~\ref{cavity}b) can simultaneously enhance the output
coupling efficiency and the radiative decay rate.  A detailed
analysis based on the finite difference time domain (FDTD) method
predicts that an optical mode volume of $0.5(\lambda/n)^3$, a $Q$
value of 30,000, a spontaneous emission decay rate enhancement
(Purcell) factor of 100, and a $\beta$-factor of 1 can be
achieved~\cite{Vuckovic03a}.  The expected photon flux in this
case is $4\times10^4$~photons/s.

\begin{figure}
\includegraphics[height = 40mm,keepaspectratio]{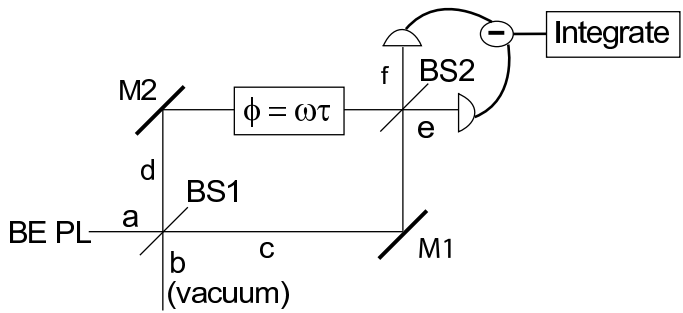}
\caption{Mach-Zender interferometer for frequency discrimination
of BE PL.} \label{directfig}
\end{figure}

In order to find the measurement time needed to determine the
nuclear spin state, we calculate the signal-to-noise  ratio for
direct frequency detection. In the scheme illustrated by
Fig.~\ref{directfig}, the BE PL is collected and sent to input $a$
of a Mach-Zender interferometer.  The signal is split by the first
beamsplitter into arms $c$ and $d$, which have a phase difference
of $\omega\tau$. Arms $c$ and $d$ are recombined at the second
beamsplitter, and the photons in outputs $e$ and $f$ are detected
by single photon detectors. The photocurrents are then subtracted,
and the resulting current is time-integrated.

%
%
%
%

Assuming an input state with mean frequency $\omega_0$ and a
Lorentzian spectral lineshape with full-width-half-max $\gamma$,
the average integrated single-photon current is proportional to
\begin{equation}
\langle{I}\rangle = e^{-\gamma\tau/{2}}\cos(\omega_0\tau).
\end{equation}
The variance of the current is proportional to
\begin{equation}
\langle\Delta{I}^2\rangle =\langle {I}^2 \rangle-\langle {I} \rangle^2 \\
= 1-e^{-\gamma\tau}\cos^2(\omega_0\tau).
\end{equation}
The two input frequencies of the BE PL, corresponding to the two
nuclear spin states, are centered around a mean frequency
$\omega_0$;  we notate these frequencies $\omega_\pm = \omega_0
\pm\Delta\omega/2,$ where $\hbar\Delta\omega=\hypsplit.$ The difference
in the integrated current between the nuclear states is the
signal, so the power signal-to-noise ratio per photon for such a
frequency detection scheme is given by
\begin{equation}
\frac{S}{N} =
    \frac{\left[
    \sin\bigl(\frac{1}{2}\Delta\omega\tau\bigr)
    -\sin\bigl(-\frac{1}{2}\Delta\omega\tau\bigr)
    \right]^2}
    {\cos^2\bigl(\frac{1}{2}\Delta\omega\tau\bigr)
    + e^{\gamma\tau}-1},
\end{equation}
in which the interferometer is biased such that $\omega_0\tau = (m
+ \frac{1}{2})\pi$. The maximum signal-to-noise ratio is 0.084 per
photon, assuming the upper bound on the PL linewidth of 150~MHz
and a delay time $\tau = 2$~ns. Assuming 100 photons per second
are collected, the minimum integration time needed for a
signal-to-noise ratio of unity is approximately 0.1~s. If we
assume an enhanced spontaneous emission decay rate with a 2D
photonic structure mentioned above and a collection efficiency of
0.5, the minimum integration time is reduced to approximately
$10^{-3}$~s.

Any practical detection scheme will suffer from detector
efficiency and dark count rates.  However, negligible dark count
rates and a detection efficiency of 0.4 can be obtained at
1.1~$\mu$m with a superconducting transition edge sensor
(TES)~\cite{Cabrera00a}. The effect of finite detector efficiency
$\ts\eta d$ is to decrease the signal-to-noise ratio by $\ts\eta
d$, extending the measurement time.

Sufficient integration of the luminescence signal measures the
nuclear spin state only if that state is stable over a sufficient
number of excitation/luminescence cycles.  Previous experiments
measuring the resonance of single nuclei have shown only
distributions over the nuclear states, as in a high-temperature
ensemble, due to the randomization of the nuclear spins during the
integration time of the measurement~\cite{Wrachtrup97a}. However,
in the case of \p\ nuclei in silicon, it is known from ESR
experiments that the equilibrium nuclear spin relaxation time
$T_1$ of isolated \p\ nuclei 
exceeds 10~hours at 1.25~K and 0.8~T~\cite{Feher59b}. Such a long
$T_1$ of solid-state nuclei is a combined consequence of the small
number of nuclear interactions and a small density of states to
which the nuclear Z\'eeman energy may be transferred. Cross
relaxation with the phosphorous-bound electron, induced by the
hyperfine interaction with the emission of a phonon, is also a
slow process.  It was measured to be 5~hours at 1.25~K and 0.8~T
and should theoretically scale to 30~seconds
at 4~K and 10~T~\cite{Feher59b}. Since the P$^0$ ground state only
lasts nanoseconds before each optical re-excitation in this
scheme, these already long equilibrium relaxation times are
negligible. The nucleus may only be appreciably destabilized by
the rapid optical excitations we introduce.

We first argue that nuclear spin relaxation due to optically
excited, delocalized conduction electrons is not expected to be a
significant effect. The number of conduction electrons required
for neutralization of the Auger-ionized donors in $1\ns$ is
approximately $10^{13}\percc,$ assuming a 4~K electron capture
cross section of $4\times 10^{-11}\cm^2$~\cite{Norton73a}. At this
temperature, electrons are expected to thermalize quickly, in
which case the $T_1$ theory described in Ref.~\cite{Abragam61a} is
applicable.  This theory predicts a $T_1$ greater than $10^{6}$
seconds. Even \si\ in heavily doped ($>10^{17}$~donors/cm$^3$)
silicon has a measured $T_1$ in excess of 200 minutes at
4~K~\cite{Sundfors64a,Jerome68a}; the $T_1$ corresponding to \p\
nuclei would only be shorter by a factor of
$\gamma_{\si}^2/\gamma_{\p}^2=4,$ still leaving this timescale
unimportant. An argument that $T_1$ due to free excitons should
also be negligibly long follows a similar reasoning, since only
the spin of the $s$-like electron interacts appreciably with the
nucleus, and Boltzmann statistics may still be assumed.

Of greater concern is the probability of a nuclear spin flip
during the capture of a free electron following the Auger process.
Such a nuclear flip arises in second order perturbation theory, in
which a virtual electron capture and a virtual electron/nuclear
spin flip-flop in the neutral-donor state occur concurrently with
the energy compensated by the real emitted phonon.  To estimate
the probability of such an event, we note that the energy cost of
a hyperfine-induced electron/nuclear spin flip-flop (approximately
the 1.2~meV Z\'eeman energy of the neutral-donor spin) is
substantially smaller than the donor binding energy (45~meV).  We
may therefore assume that the density-of-states factors in Fermi's
Golden Rule are unchanged between the first and second order
processes, and that they are independent of the initial spin
state. We also assume that the optical excitation of free carriers
is not spin selective, and since $T_1$ for these carriers exceeds
the capture time, this implies that the initial spin polarizations
are approximately equal. It follows from these assumptions that
the probability ratio between the first and second order processes
is well approximated by the ratio between their matrix elements.
The second order matrix element for an electron/nuclear flip-flop
process, assuming without loss of generality that the nucleus
begins in the $\ket{\Downarrow}$ state, may be written as
\begin{equation}
\left|\secondorder\right|^2=  \left| \frac{
    \bra\startstate\transit\ket\middlestate
    \bra\middlestate\ham{C}\ket\badfinalstate}
    {\energydiff}
\right|^2,
\end{equation}
in which $\ket\startstate$ describes the ionized donor and
$\ket\middlestate$ describes the neutral donor with electron spin
up and nucleus spin down.  These states are the only important
ones for nuclear destabilization, since the $\ket\nonstate$ state
is unperturbed by $\ham{C}$. The unperturbed eigenenergy
difference between the intermediate and final state,
$\energydiff,$ is a sum of the electron Z\'eeman term, the nuclear
Z\'eeman term, and $\hypsplit$; the electron Z\'eeman term
dominates this sum. The Hamiltonian term $\transit$ refers to the
interaction leading to the capture of the free electron; thus, a
first-order matrix element for a neutralization process without a
flip-flop is written as
\begin{equation}
\left| \firstorderup\right|^2
    =
    \left|
    \bra\startstate\transit\ket\goodfinalstate
    \right|^2.
\end{equation}
We assume that this Coulombic process has no spin selectivity. It
follows that the probability per transition of a flip-flop may be
written as
\begin{multline}
\frac{\left|\secondorder\right|^2}
     {\left| \firstorderup\right|^2+\left| \firstorderdown\right|^2+\left| \secondorder\right|^2}
    \\ \approx\frac{1}{2}\left|
    \frac{
    \bra\middlestate
    \ham{C}\ket\badfinalstate
    }{\energydiff}
    \right|^2
    =\frac{1}{2}\left(\frac{\Delta \ts{E}{HF,P$^0$}}{{g}\mub B_0}\right)^2.
\label{prob}
\end{multline}
This probability can be seen from an alternative viewpoint: if we
use second-order time-independent perturbation theory to calculate
the mixing of the $\ket\middlestate$ and $\ket\badfinalstate$
states due to the flip-flop terms of $\ham{C},$ and presume that
capture rates to these perturbed states are the same as to the
unperturbed states, the same probability is obtained.

The free exciton capture process leads to a similar second-order
probability for nuclear randomization. In this case, however, the
binding energy of the free exciton to the neutral donor is only
somewhat larger than the electron Z\'eeman energy, so the
density-of-states factors in the transition rates can become
important.  However, we do not expect this correction to alter the
per-transition probability by more than a factor of order unity.
We thus estimate that at 10~T, the probability of a nuclear flip
for each free-exciton-capture and free-electron-capture process is
approximately twice the result of Eq. (\ref{prob}):
(60~MHz/280~GHz)$^2=5\times 10^{-8}$. The consequences of this
probability will be discussed shortly.

A nuclear spin flip can also occur in a similar manner during a
free exciton capture or BE Auger decay due to the BE
(hole/nuclear) hyperfine interaction. This process is less
important because the BE hyperfine interaction is much weaker. The
magnitude of this interaction, which includes both dipolar
coupling and any contribution due to a small $s$-like component of
the hole's Bloch wavefunction, may be estimated as 2~MHz from the
results of muon spin resonance experiments~\cite{Mamedov02a}.
Since the BE Z\'eeman energy is comparable to that of the bound
electron, the probability of a spin flip due to the BE hyperfine
coupling is smaller by the approximate factor
$(2\MHz/60\MHz)^2\sim 10^{-3}.$ Nuclear spin flip due to the P$^0$
hyperfine coupling during radiative decay has a similar order to
that during free electron capture, but since radiative decay is
7000 times less frequent, this probability may also be neglected.

It is unfortunate that the predominant decay mechanism is
non-radiative, since each Auger process increases the probability
of nuclear randomization without providing a signal photon. After
$2\times10^6$ excitations, at which point the probability of
nuclear randomization exceeds 1/10, we can only expect to have
collected and detected $25$~photons or $5\times10^3$ photons with
a DBR planar cavity or a 2D photonic crystal, respectively. This
still yields a usable signal-to-noise ratio of 2 or $400$ for the
respective geometries. Thus, we expect the measurement-induced
lifetime of the single nuclear spin to be long enough to allow for
the measurement of the nuclear spin {state}.

In summary, we have proposed utilizing the sharp BE PL lines for
single nuclear spin measurement.  We have demonstrated that it is
theoretically possible to measure a single donor impurity spin in
silicon using currently available technology -- a semiconductor
microcavity and single-photon detectors. We have analyzed only one
type of BE in Si, due to the experimentally observed strong
hyperfine coupling and narrow PL linewidths. However, bound
excitons exist in abundance in most semiconductors. Thus, this
technique could be applied to many more systems if the crystal
quality is sufficiently high and the hyperfine coupling and photon
flux are sufficiently large.  {We believe this technique holds
promise for state readout in quantum computers utilizing nuclear
qubits in semiconductors}.

This work was partially supported by NTT Basic Research
Laboratories.  T. D. L. was supported by the Fannie and John Hertz
Foundation.  We would like to thank J. Vuckovic for fruitful
discussions.

\end{document}